\documentclass[pra,twocolumn,floatfix,a4paper,superscriptaddress]{revtex4-2}
\usepackage{bm,color,graphicx,amsmath,txfonts}

\usepackage{hyperref}
\usepackage{chngcntr}
\usepackage{textcomp}

\hypersetup{
 colorlinks=true,   
 linkcolor=blue,  
 citecolor=blue, 
 urlcolor =blue    
}

\begin{document}

\title{Strong squeezing of microwave output fields via reservoir-engineered cavity magnomechanics}

\author{Hang Qian}
\affiliation{Interdisciplinary Center of Quantum Information, State Key Laboratory of Extreme Photonics and Instrumentation, and Zhejiang Province Key Laboratory of Quantum Technology and Device, Department of Physics, Zhejiang University, Hangzhou 310027, China}
\author{Xuan Zuo}
\affiliation{Interdisciplinary Center of Quantum Information, State Key Laboratory of Extreme Photonics and Instrumentation, and Zhejiang Province Key Laboratory of Quantum Technology and Device, Department of Physics, Zhejiang University, Hangzhou 310027, China}
\author{Zhi-Yuan Fan}
\affiliation{Interdisciplinary Center of Quantum Information, State Key Laboratory of Extreme Photonics and Instrumentation, and Zhejiang Province Key Laboratory of Quantum Technology and Device, Department of Physics, Zhejiang University, Hangzhou 310027, China}
\author{Jiong Cheng}
\affiliation{Interdisciplinary Center of Quantum Information, State Key Laboratory of Extreme Photonics and Instrumentation, and Zhejiang Province Key Laboratory of Quantum Technology and Device, Department of Physics, Zhejiang University, Hangzhou 310027, China}
\affiliation{Department of Physics, School of Physical Science and Technology, Ningbo University, Ningbo, 315211, China}
\author{Jie Li}\thanks{jieli007@zju.edu.cn}
\affiliation{Interdisciplinary Center of Quantum Information, State Key Laboratory of Extreme Photonics and Instrumentation, and Zhejiang Province Key Laboratory of Quantum Technology and Device, Department of Physics, Zhejiang University, Hangzhou 310027, China}

\begin{abstract}
We show how to achieve strong squeezing of a microwave output field by reservoir engineering a cavity magnomechanical system, consisting of a microwave cavity, a magnon mode, and a mechanical vibration mode.  The magnon mode is simultaneously driven by two microwave fields at the blue and red sidebands associated with the vibration mode. The two-tone drive induces a squeezed magnonic reservoir for the intracavity field, leading to a squeezed cavity mode due to the cavity-magnon state swapping, which further yields a squeezed cavity output field. The squeezing of the output field is stationary and substantial using currently available parameters in cavity magnomechanics. The work indicates the potential of the cavity magnomechanical system in preparing squeezed microwave fields, and may find promising applications in quantum information science and quantum metrology.
\end{abstract}

\maketitle

\section{Introduction}

The past decade has witnessed fast developments in the field of cavity magnonics based on collective spin excitations (magnons) in ferrimagnetic materials, such as yttrium iron garnet (YIG). As one of the prominent advantages, ferrimagnetic magnons show an excellent ability to coherently interact with many different quantum systems, such as microwave/optical photons, vibration phonons and superconducting qubits~\cite{APE,NRM,Yuan,Gerrit,Zuo}. The coupling to vibration phonons by magnetostriction forms the system of cavity magnomechanics (CMM), which studies the interactions among microwave cavity photons, magnons, and phonons~\cite{Zuo,Tang,Jie18,Davis,Li22,Li23}. The CMM system promises diverse potential applications in quantum information processing, quantum metrology, and the study of macroscopic quantum states~\cite{Zuo,Jie18,Jie20,nsr,LPR}.    So far, magnomechanically induced transparency and absorption~\cite{Tang}, mechanical cooling and amplification~\cite{Davis,Li22}, magnon-phonon cross-Kerr effect and mechanical bistability~\cite{Li22}, and magnonic frequency combs~\cite{Dong23} have been experimentally demonstrated. Very recently, the CMM system has realized the strong coupling and achieved a high cooperativity of $4\times 10^3$~\cite{Li23}, which enables coherent quantum control of magnons, photons, and phonons. Theoretical studies have been focused on the preparation of various quantum states~\cite{Jie18,Jie20,nsr,Jie19a,Jie19b,Tan,Ding,QST,HFW,Wuj,Irfan,Nie,qst2,qst3,FR23,WHF22,Ding22,Asjad,Fan23a,LPR}, and the applications of the system in quantum information science, quantum technologies, and high-precision measurements, including slow light~\cite{Xiong,WuZ,JH2}, phonon laser~\cite{Ding2,WangC}, thermometry~\cite{Davis2}, quantum memory~\cite{JJ,Twamley}, exceptional points~\cite{JH},  frequency combs~\cite{XH,XH2}, high-order sidebands~\cite{Liu19,XH3}, etc.

In this article, we propose to generate significant squeezing of the microwave output field in the CMM system by exploiting the reservoir engineering. The idea of reservoir engineering was initially proposed for the trapped-ion system~\cite{RE1,RE2}. Its subsequent application to the optomechanical system~\cite{RE3,RE4,RE5,RE7,RE8,RE10} led to successful demonstrations of the squeezing~\cite{sqz1,sqz2,sqz3} and entanglement~\cite{en1} of massive mechanical resonators.  Specifically, in the present CMM system, the magnon mode is simultaneously driven by two microwave fields that are resonant with the blue and red sidebands associated with the mechanical mode. The two-tone drive leads to the squeezing of both the mechanical and magnon modes. Consequently, the cavity mode gets squeezed due to the electromagnonic state-swap interaction, thus yielding a squeezed cavity output field.  We provide optimal conditions for achieving significantly large squeezing of the microwave output field.
We would like to note that the reservoir engineering has been applied to the CMM system to obtain strong squeezing of the mechanical mode~\cite{HFW}. Although intracavity squeezing has also been studied, the squeezing is small, nonstationary, and inaccessible~\cite{HFW}. Here, we aim to achieve strong and stationary squeezing of {\it microwave output fields}, which can be directly accessed and applied to quantum information science and quantum metrology.

The article is organized as follows. In Sec.~\ref{model}, we introduce the model and provide the Hamiltonian and Langevin equations of the system. In Sec.~\ref{NSD}, we derive the noise spectrum density (NSD) of the cavity output field and show the results of the output field squeezing under the rotating-wave approximation (RWA). We further analyze the squeezing performance and provide optimal conditions for achieving strong squeezing. In Sec.~\ref{exact}, we solve the full dynamics by including the counter-rotating-wave (CRW) terms and study their impact on the squeezing. Finally, we summarize our findings in Sec.~\ref{conc}.

\section{The system}
\label{model}

\begin{figure}[t]
\hskip0.0cm\includegraphics[width=0.85\linewidth]{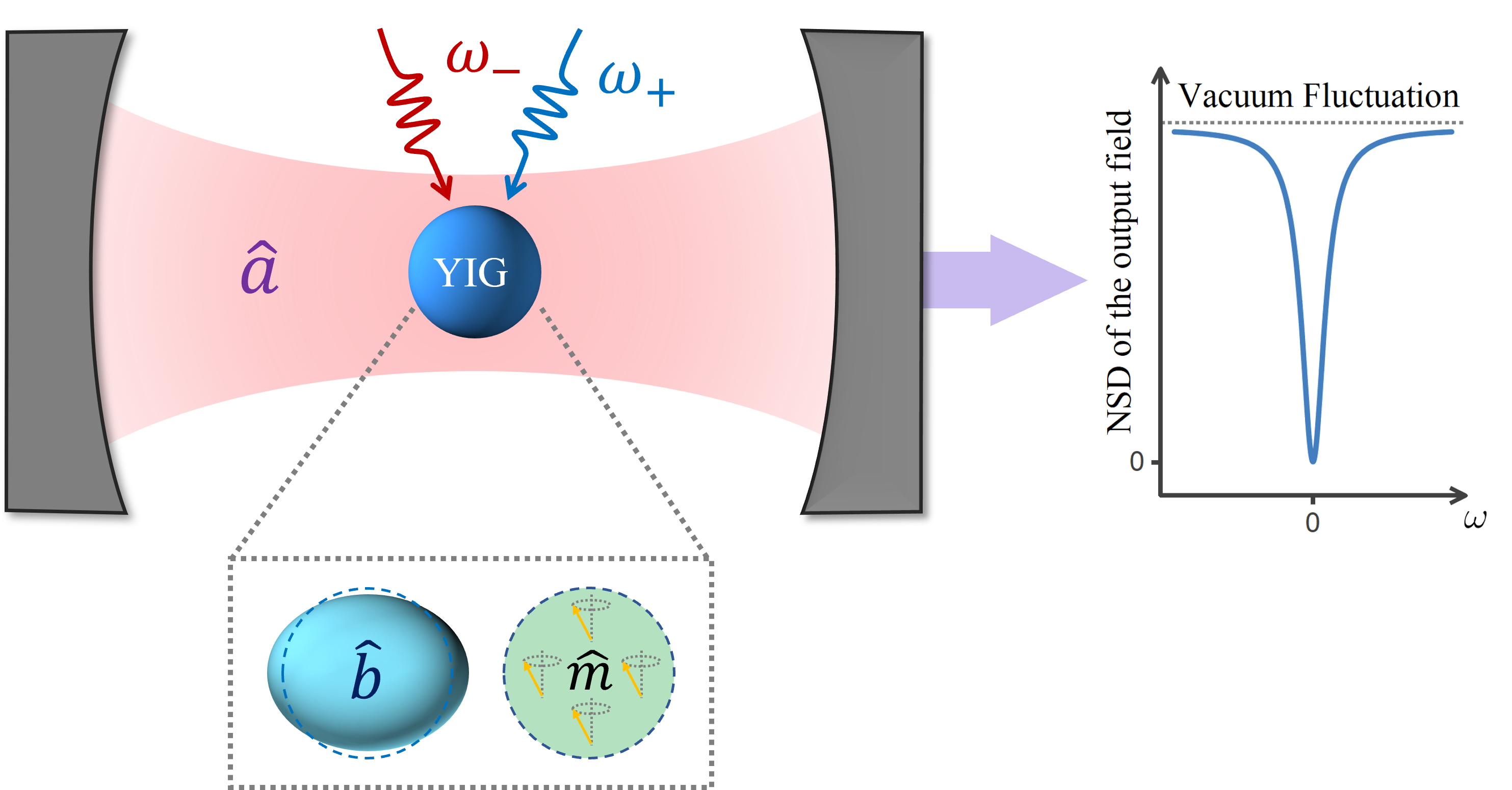}
\caption{Sketch of the reservoir-engineered CMM system used to achieve strong squeezing of the microwave cavity output field. The reservoir engineering is implemented by simultaneously driving the magnon mode with two microwave fields at the frequencies $\omega_{\pm}$.}
\label{fig1}
\end{figure}

The CMM system consists of a microwave cavity mode, a magnon mode, and a mechanical vibration mode; see Fig.~\ref{fig1}. The magnon mode represents a collective motion of a large number of spins in a ferrimagnet, e.g., a YIG sphere~\cite{Tang,Jie18,Davis,Li22,Li23}, which couples to the microwave cavity via the magnetic-dipole interaction and to the vibration mode of the YIG sphere via the magnetostrictive interaction. We consider a large-size YIG sphere with the diameter in the range from $10^2$ to $10^3$ $\mu$m~\cite{Tang,Jie18,Davis,Li22,Li23}, such that the resonance frequency of the mechanical mode is much lower than that of the magnon mode, promising a radiation-pressure-like dispersive coupling between magnons and phonons~\cite{qst2}. The magnon mode is simultaneously driven by two microwave fields loaded, e.g., via a loop antenna~\cite{Li23}.  The Hamiltonian of the system reads
\begin{equation}\label{Hamiltonian}
\begin{split}
   \hat{H}/\hbar&=\omega_a \hat{a}^\dagger \hat{a} +\omega_m \hat{m}^\dagger \hat{m} +\omega_b \hat{b}^\dagger \hat{b} +g\left(\hat{a}^\dagger \hat{m}+\hat{a}\hat{m}^\dagger\right)\\
   &+G_0\hat{m}^\dagger \hat{m}\left( \hat{b} + \hat{b}^\dagger \right)+\left[ \left(\Omega_+e^{-i\omega_+ t}+\Omega_-e^{-i\omega_-t}\right)\hat{m}^\dagger+{\rm h.c.} \right],
\end{split}
\end{equation}
where $\hat{a}$, $\hat{m}$ and $\hat{b}$ ($\hat{a}^\dagger$, $\hat{m}^\dagger$ and $\hat{b}^\dagger$) are the annihilation (creation) operators of the cavity, magnon, and mechanical modes, respectively, and $\omega_j$ ($j=a$, $m$, $b$) are the resonance frequencies of the three modes. $g$ is the cavity-magnon coupling strength and $G_0$ denotes the bare magnomechanical coupling strength, which is typically weak, but the effective magnomechanical coupling $G$ can be significantly improved by applying a strong drive field to the magnon mode.   Here, the magnon mode is driven by two microwave fields at the frequencies $\omega_{\pm}=\omega_m\pm\omega_b$, with the associated magnon-drive coupling strengths $\Omega_{\pm}$ (i.e., the Rabi frequencies~\cite{Jie18}).   

The quantum Langevin equations (QLEs) of the system can be obtained by including the dissipation and input noise of each mode, given by
\begin{equation}\label{QLEs_all}
  \begin{split}
     \dot{\hat{a}} =  & -\left(i\omega_a+\frac{\kappa_a}{2}\right)\hat{a}-ig\hat{m}+\sqrt{\kappa_a}\hat{a}_{in} ,\\
       \dot{\hat{b}}=&-\left( i\omega_b+\frac{\gamma_b}{2}\right) \hat{b}-iG_0\hat{m}^\dagger \hat{m}+\sqrt{\gamma_b}\hat{b}_{in}, \\
      \dot{\hat{m}}  =&-\left( i\omega_m+\frac{\kappa_m}{2}\right) \hat{m}-ig\hat{a}-iG_0\hat{m}\left( \hat{b}+\hat{b}^\dagger\right)\\
      & -i\left(\Omega_+  e^{-i\omega_+t}+\Omega_-e^{-i\omega_-t}\right)+\sqrt{\kappa_m}\hat{m}_{in},
  \end{split}
\end{equation}
where $\kappa_a$, $\gamma_b$ and $\kappa_m$ are the dissipation rates of the cavity, mechanical, and magnon modes, respectively, and $\hat{O}_{in}$ ($O=a,b,m$) denote the input noises of the three modes, of which the non-zero correlation functions are $\langle\hat{O}_{in}^\dagger(t)\hat{O}_{in}(t')\rangle = N_O\left(\omega_O\right) \delta\left(t-t'\right)$ and $\langle\hat{O}_{in}(t)\hat{O}_{in}^ \dagger(t')\rangle=\left[N_O(\omega_O)+1\right] \delta(t-t')$, with $N_O(\omega_O)=\left[{\rm exp}(\frac{\hbar\omega_O}{k_BT})-1\right]^{-1}$ being the equilibrium mean thermal excitation number of each mode and $T$ being the bath temperature.

Following the standard linearization treatment, we write each mode operator $\hat{O}$ as a large classical average $O_s$ plus a small fluctuation operator $\delta\hat{O}$, i.e., $\hat{O}=O_s +\delta\hat{O}$ ($O=a,b,m$). Substituting them into Eq.~\eqref{QLEs_all}, the equations are then separated into two sets of equations for classical averages and quantum fluctuations, respectively.  Since the amplitudes of the magnon and cavity modes are dominant at the two drive frequencies $\omega_{\pm}$, we assume that $m_s \simeq m_+ e^{-i\omega_+t}+m_-e^{-i\omega_-t}$ and $a_s \simeq a_+e^{-i\omega_+t}+a_-e^{-i\omega_-t}$~\cite{RE3,RE4,RE5,RE7,RE8}. Because of the very small bare coupling $G_0$ in current CMM experiments~\cite{Tang,Davis,Li22,Li23}, typically $G_0 |{\rm Re} \,b_s| \ll \omega_m$. This allows us to safely neglect the small term $iG_0m_s(b_s+b_s^*)$ in solving the set of equations for the classical averages.  We therefore obtain the following mean amplitudes $m_{\pm}$ associated with the two frequency components of the magnon mode:
\begin{equation}\label{ms}
  m_{\pm}=\frac{\Omega_{\pm}}{\omega_{\pm}-\omega_m+i\frac{\kappa_m}{2}
  -\frac{g^2}{\omega_{\pm}-\omega_a+i\frac{\kappa_a}{2}}}.
\end{equation}

By neglecting small second-order fluctuation terms, we obtain the linearized QLEs for the fluctuations, which, in the interaction picture with respect to $\hbar\omega_0 \left(\hat{a}^\dagger \hat{a} + \hat{m}^\dagger \hat{m} \right) + \hbar\omega_b \hat{b}^\dagger \hat{b}$, are given by
\begin{equation}\label{QLEs_fluctuation}
  \begin{split}
     \delta\dot{\hat{a}} = & -\frac{\kappa_a}{2} \delta\hat{a} - ig\delta\hat{m}+\sqrt{\kappa_a} \hat{a}_{in}, \\
     \delta\dot{\hat{b}} = & -\frac{\gamma_b}{2} \delta\hat{b} - i\left(G_-+G_+e^{2i\omega_bt}\right) \delta\hat{m}, \\ &\quad\quad\quad-i\left(G_++G_-e^{2i\omega_bt}\right)\delta\hat{m}^\dagger + \sqrt{\gamma_b}\hat{b}_{in}, \\
     \delta\dot{\hat{m}} = & - \frac{\kappa_m}{2}\delta\hat{m} -ig\delta\hat{a}  - i\left(G_-+G_+e^{-2i\omega_bt}\right)\delta\hat{b}\\
      &\quad \quad \quad - i\left(G_++G_-e^{2i\omega_bt}\right)\delta\hat{b}^\dagger +\sqrt{\kappa_m} \hat{m}_{in}.
  \end{split}
\end{equation}
We consider the resonant case of $\omega_a=\omega_m \equiv \omega_0$, and $G_{\pm}=G_{0}m_{\pm}$ are the effective magnomechanical coupling strengths associated with the two drive fields, which are generally complex, but can be set real by adjusting the phases of the drive fields to have real $m_{\pm}$ [cf. Eq.~\eqref{ms}].  Without loss of generality, we consider real and positive couplings $G_{\pm} >0$ and $G_->G_+$ for keeping the system being stable~\cite{RE3,RE4,RE5,RE7,RE8,RE10}.

\section{Squeezing of the cavity output field under the RWA}
\label{NSD}

It is difficult to solve Eq.~\eqref{QLEs_fluctuation} analytically due to the non-resonant time-dependent terms. However, it is doable by taking the RWA and neglecting the CRW terms under the condition of $\kappa_{a(m)}, g, G_{\pm} \ll \omega_b$~\cite{RE7}, which leads to the following QLEs:
\begin{equation}\label{QLEs_RWA}
  \begin{split}
     \delta\dot{\hat{a}} & = - \frac{\kappa_a}{2} \delta \hat{a} - ig\delta\hat{m} + \sqrt{\kappa_a} \hat{a}_{in}, \\
     \delta\dot{\hat{b}} & = -\frac{\gamma_b}{2} \delta \hat{b} - i \left( G_{-}\delta\hat{m} + G_{+} \delta\hat{m}^\dagger \right) + \sqrt{\gamma_b} \hat{b}_{in}, \\
     \delta\dot{\hat{m}}  & = - \frac{\kappa_m}{2}\delta\hat{m} -ig\delta\hat{a}  -i \left( G_{-} \delta\hat{b} +G_{+}\delta\hat{b}^\dagger \right) + \sqrt{\kappa_m} \hat{m}_{in}.
  \end{split}
\end{equation}
 The above equations can be conveniently solved in the frequency domain by taking the Fourier transforms $O[\omega] = \int_{-\infty}^{\infty} O(t)e^{i\omega t}\, \mathrm{d}t$ and $O^\dagger[-\omega] =\left( \int_{-\infty}^{\infty} O(t)e^{-i\omega t}\, \mathrm{d}t \right)^\dagger = \int_{-\infty}^{\infty} O^\dagger(t)e^{i\omega t} \,\mathrm{d}t$, and we obtain
\begin{widetext}
\begin{equation}\label{cavity field}
  \begin{split}
     \delta\hat{a}[\omega] & = \chi_a^{\rm eff} [\omega] \left[ \sqrt{\kappa_a} \hat{a}_{in}[\omega] - i g \chi_{\rm mb}[\omega] \sqrt{\kappa_m}  \hat{m}_{in} [\omega] - g \chi_{\rm mb}[\omega] \sqrt{\gamma_b}  \chi_b [\omega] \left(G_{-} \hat{b}_{in}[\omega] + G_+ \hat{b}_{in}^\dagger [-\omega] \right) \right],\\
     \delta\hat{b} [\omega] & = \chi_b^{\rm eff} [\omega] \left[ \sqrt{\gamma_b} \hat{b}_{in} [\omega] - i \chi_{\rm ma}[\omega] \sqrt{\kappa_m} \left(G_- \hat{m}_{in} [\omega] +G_+m_{in}^\dagger [-\omega] \right) - g \chi_{\rm ma}[\omega]  \chi_a[\omega] \sqrt{\kappa_a} \left(G_- \hat{a}_{in} [\omega] -G_+ \hat{a}_{in}^\dagger [-\omega] \right)\right], \\
     \delta\hat{m} [\omega] & = \chi_m^{\rm eff} [\omega] \left[ \sqrt{\kappa_m} \hat{m}_{in} [\omega] -i g\chi_a[\omega] \sqrt{\kappa_a} a_{in} [\omega] - i \chi_b[\omega] \sqrt{\gamma_b}  \left( G_-\hat{b}_{in} [\omega] +G_+ \hat{b}_{in} ^\dagger [-\omega] \right)  \right],
  \end{split}
\end{equation}
\end{widetext}
where $ \chi_j^{\rm eff}[\omega]$ ($j=a,b,m$) are the effective susceptibilities of the cavity, mechanical, and magnon modes, respectively, i.e.,
\begin{equation}\label{chi_eff}
  \begin{split}
     \chi_a^{\rm eff}[\omega] & = \left( \chi_a^{-1}[\omega] + g^2 \chi_{\rm mb} [\omega] \right)^{-1}, \\
     \chi_b^{\rm eff} [\omega] & = \left( \chi_b^{-1}[\omega] + \tilde{G}^2 \chi_{\rm ma} [\omega]\right)^{-1}, \\
     \chi_m^{\rm eff} [\omega] & = \left( \chi_m^{-1} [\omega] + g^2 \chi_a [\omega] + \tilde{G}^2 \chi_b[\omega]\right)^{-1},
  \end{split}
\end{equation}
which are defined by means of  the natural susceptibilities $\chi_j [\omega] = \left(\frac{\kappa_j}{2} - i\omega\right)^{-1}$ ($\kappa_j = \kappa_a, \gamma_b, \kappa_m$), and the complex susceptibilities $\chi_{\rm mb} [\omega] = \left( \chi_m^{-1}[\omega] + \tilde{G}^2 \chi_b[\omega] \right)^{-1} $ and $\chi_{\rm ma} [\omega] = \left( \chi_m^{-1} [\omega] +g^2\chi_a [\omega]\right)^{-1}$, with $\tilde{G}=\sqrt{G_-^2-G_+^2}$.

We aim to obtain squeezed microwave output fields (rather than unaccessible intracavity fields), and thus study the noise property of the cavity output field. The fluctuation of the cavity output field $\delta \hat{a}_{\rm out}$ can be achieved using the standard input-output relation, $\delta \hat{a}_{\rm out} [\omega] = \sqrt{\kappa_a}\delta \hat{a} [\omega] - \hat{a}_{in}[\omega]$~\cite{GC}. We can then define the fluctuation of the general quadrature of the output field as
\begin{equation}\label{general quaduature}
  \delta U_{\rm out}[\omega,\phi]=\frac{1}{\sqrt{2}} \left[ \delta \hat{a}_{\rm out} [\omega] e^{-i\phi} + \delta \hat{a} _{\rm out}^\dagger [-\omega] e^{i\phi}\right],
\end{equation}
with $\phi$ being the phase angle, and the NSD of the output field is thus obtained by
\begin{equation}\label{NSD_def}
  \begin{split}
     S^{\rm out}_{\phi}[\omega] = \frac{1}{4\pi} &
     \int_{-\infty}^{\infty} \mathrm{d\omega'}\, e^{-i(\omega+\omega')t} \times \\
       &\left \langle \delta U_{\rm out} [\omega] \delta U_{\rm out} [\omega'] + \delta U_{\rm out} [\omega'] \delta U_{\rm out} [\omega] \right\rangle.
  \end{split}
\end{equation}
After some calculation, we obtain the following analytical form of the NSD of the cavity output field:
\begin{equation}\label{NSD_result}
  S_{\phi}^{\rm out} [\omega]=S^{a} [\omega] + S^{m} [\omega] +S_{\phi}^{b}[\omega] ,
\end{equation}
where
\begin{equation}\label{NSD_seperate}
  \begin{split}
     S^{a} [\omega] & = \left| \kappa_a\chi_a^{\rm eff} [\omega] -1\right|^2 \left(N_a + \frac{1}{2}\right), \\
     S^{m} [\omega] & = \kappa_a \kappa_m \left| \chi_a^{\rm eff} [\omega] g \chi_{\rm mb} [\omega] \right| ^2 \left( N_m+\frac{1}{2}\right), \\
     S_{\phi}^{b} [\omega] & =\kappa_a\gamma_b \left|\chi_a^{\rm eff}[\omega] g \chi_{\rm mb}[\omega] \chi_b[\omega]\right|^2 \times \\
     & \quad \left[ G_-^2+G_+^2 +2G_-G_+ \cos(2\phi)\right] \left(N_b+\frac{1}{2}\right)
  \end{split}
\end{equation}
characterize three different sources of noise from the cavity, magnon, and mechanical modes, respectively, and the sum of them determines the noise property of the output field.

\begin{figure}[b]
\hskip0.0cm\includegraphics[width=0.9\linewidth]{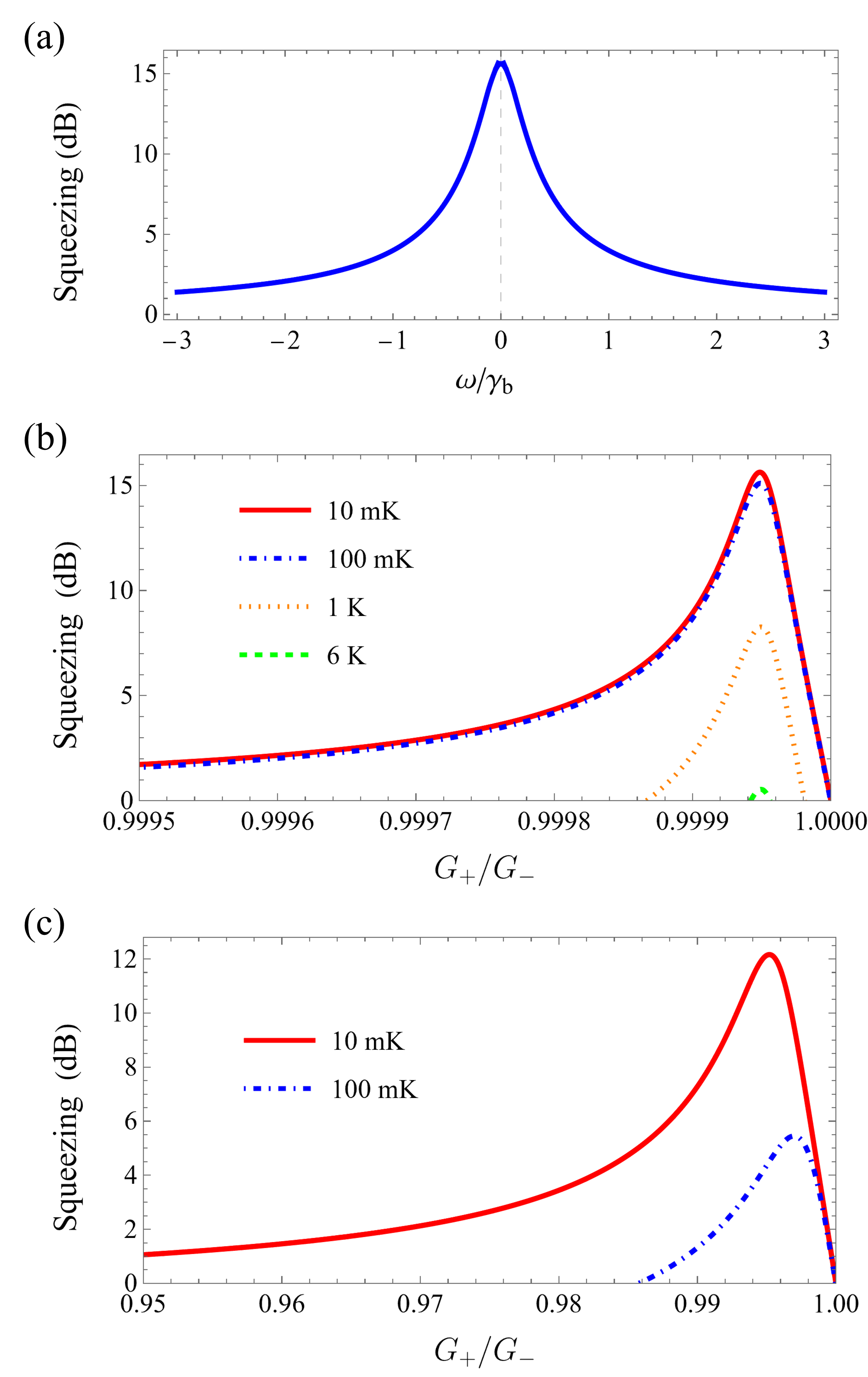}
\caption{The degree of squeezing $S$ of the cavity output field versus (a) frequency $\omega$; (b) and (c) $G_+/G_-$ for various bath temperatures. We take $T=10$ mK in (a) and $\omega=0$ in (b) and (c), and $G_+$ is optimized for the squeezing in (a). The mechanical damping rate $\gamma_b/2\pi=100$ Hz in (a) and (b), and $\gamma_b/2\pi  =10$ kHz in (c). See text for the other parameters. }
\label{fig2}
\end{figure}

It is clear from Eq.~\eqref{NSD_seperate} that an optimal squeezing angle $\phi=\frac{\pi}{2}$ corresponds to a minimum noise of $S^{b} [\omega]$ from the mechanical mode, since the couplings are assumed to be positive, $G_{\pm}>0$.  In Fig.~\ref{fig2}(a), we plot the NSD of the output field $S_{\phi}^{\rm out}[\omega]$ at $\phi=\frac{\pi}{2}$ and the associated degree of squeezing $S$ (in units of dB), i.e., $S = -10\rm{log}_{10}[{\it S}_{\pi/2}^{\rm out}/{\it S}_{\rm vac}]$, where $S_{\rm vac}=\frac{1}{2}$ corresponds to the vacuum noise. We adopt experimentally feasible parameters~\cite{Tang,Davis,Li22,Li23}: $\omega_a/2\pi=\omega_m/2\pi=10$~GHz, $\omega_b/2\pi=30$~MHz, $\gamma_b/2\pi=100$~Hz, $\kappa_a/2\pi=\kappa_m/2\pi=1$~MHz, $T=10$~mK, $g/2\pi=3$~MHz, and $G_-/2\pi=3$~MHz, satisfying the condition of $\kappa_{a(m)}, g, G_{\pm} \ll \omega_b$ for taking the RWA. $G_+$ is a free parameter to optimize the squeezing, which can be easily adjusted by varying the strength of the drive field with frequency $\omega_+$.  The results show that optimal squeezing is achieved at the cavity resonance $\omega=0$~\cite{AK14}.

At the optimal conditions of $\phi=\frac{\pi}{2}$ and $\omega=0$, the NSD of the output field takes the following simple form:
\begin{equation}\label{Sout}
\begin{split}
  S_{\pi/2}^{\rm out} [0] &= \left(1- 8 g^2\gamma_b \Xi  \right)^2 \Big(N_a + \frac{1}{2}\Big) + 16 \kappa_a \kappa_m g^2 \gamma_b^2  \Xi^2 \Big( N_m+\frac{1}{2}\Big)  \\
  &+ 64 \kappa_a\gamma_b  g^2 \Xi^2 \Big(G_- -G_+ \Big)^2 \Big(N_b+\frac{1}{2}\Big),
 \end{split}
\end{equation}
where $\Xi^{-1}=4\tilde{G}^2\kappa_a + 4g^2\gamma_b + \kappa_a\gamma_b\kappa_m$. Significant squeezing of the output field can be achieved when the coefficients in the above three noise terms approach zero.

\begin{figure}[b]
\hskip0.0cm\includegraphics[width=\linewidth]{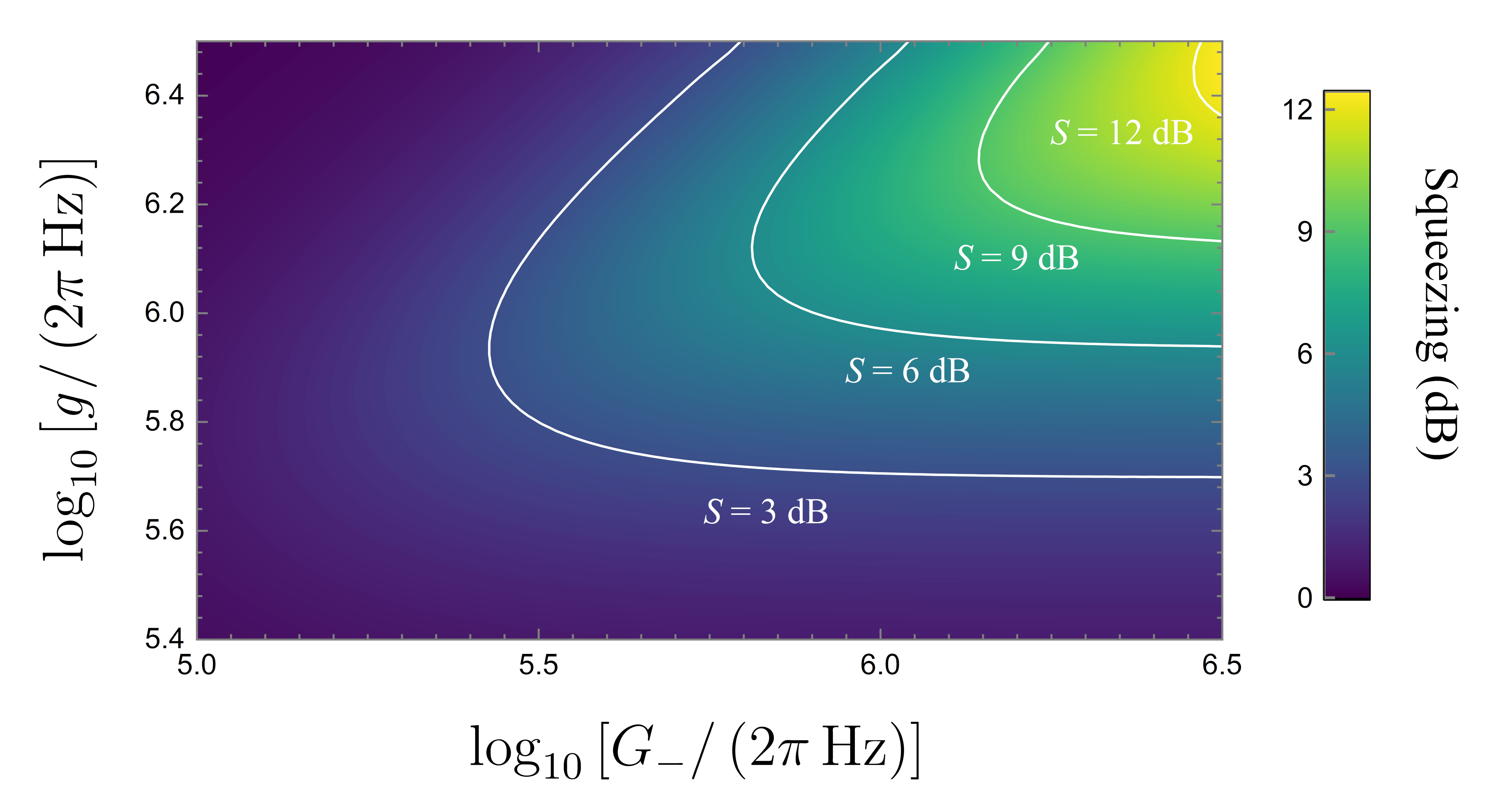}
\caption{Density plot of the output field squeezing versus couplings $G_-$ and $g$. We take $T=10$ mK, and $G_+$ is optimized. The other parameters are the same as those in Fig.~\ref{fig2}(c).}
\label{fig3}
\end{figure}

Figure~\ref{fig2}(b) shows the squeezing of the output field as a function of the ratio $G_+/G_-$ for a fixed coupling $G_-$. Clearly, there is an optimal ratio $G_+/G_-$ for a given temperature, as a result of the trade-off between the cooling and squeezing interactions associated with the two drive fields~\cite{RE3,RE4,RE5,RE7,RE8,RE10}.   As the temperature rises, the output field squeezing degrades and the optimal ratio $G_+/G_-$ {\it slightly} increases. This trend of the optimal $G_+/G_-$ as the temperature increases is consistent with that for the intracavity squeezing, but is opposite to that for the mechanical squeezing. This can be seen by comparing with the results of the intracavity and mechanical squeezing provided in the Appendix. The squeezing is quite robust against bath temperature, and $S>0$ for the temperature being up to $\sim6$~K. 
It should be noted that a high precision on $G_{+}/G_{-}$ is required to achieve the optimal squeezing in Fig.~\ref{fig2}(b), which is very challenging in the experiment.  The high requirement on the precision can, however, be relaxed by adding more noises into the system~\cite{RE10}, e.g., by increasing the mechanical damping rate, at the price of the reduction of the degree of squeezing. Figure~\ref{fig2}(c) is obtained with a larger mechanical damping rate $\gamma_b/2\pi  =10$ kHz, which can be realized by increasing the contact area between the YIG sphere and the supporting fiber~\cite{Tang}.  Clearly, although the maximum degree of squeezing is reduced from ${\sim}16$ dB to ${\sim}12$ dB, the precision on  $G_{+}/G_{-}$ is much relaxed.  With the precision of $0.1\%$ on the drive power for the analog signal generator (Keysight PSG E8257D) used in the experiments~\cite{Li22,Li23}, corresponding to the precision of $0.15\%$ on $G_{+}/G_{-}$ under the parameters of Fig.~\ref{fig2}, the optimal squeezing in Fig.~\ref{fig2}(c) can be practically achieved.

\begin{figure}[t]
\hskip0.0cm\includegraphics[width=0.95\linewidth]{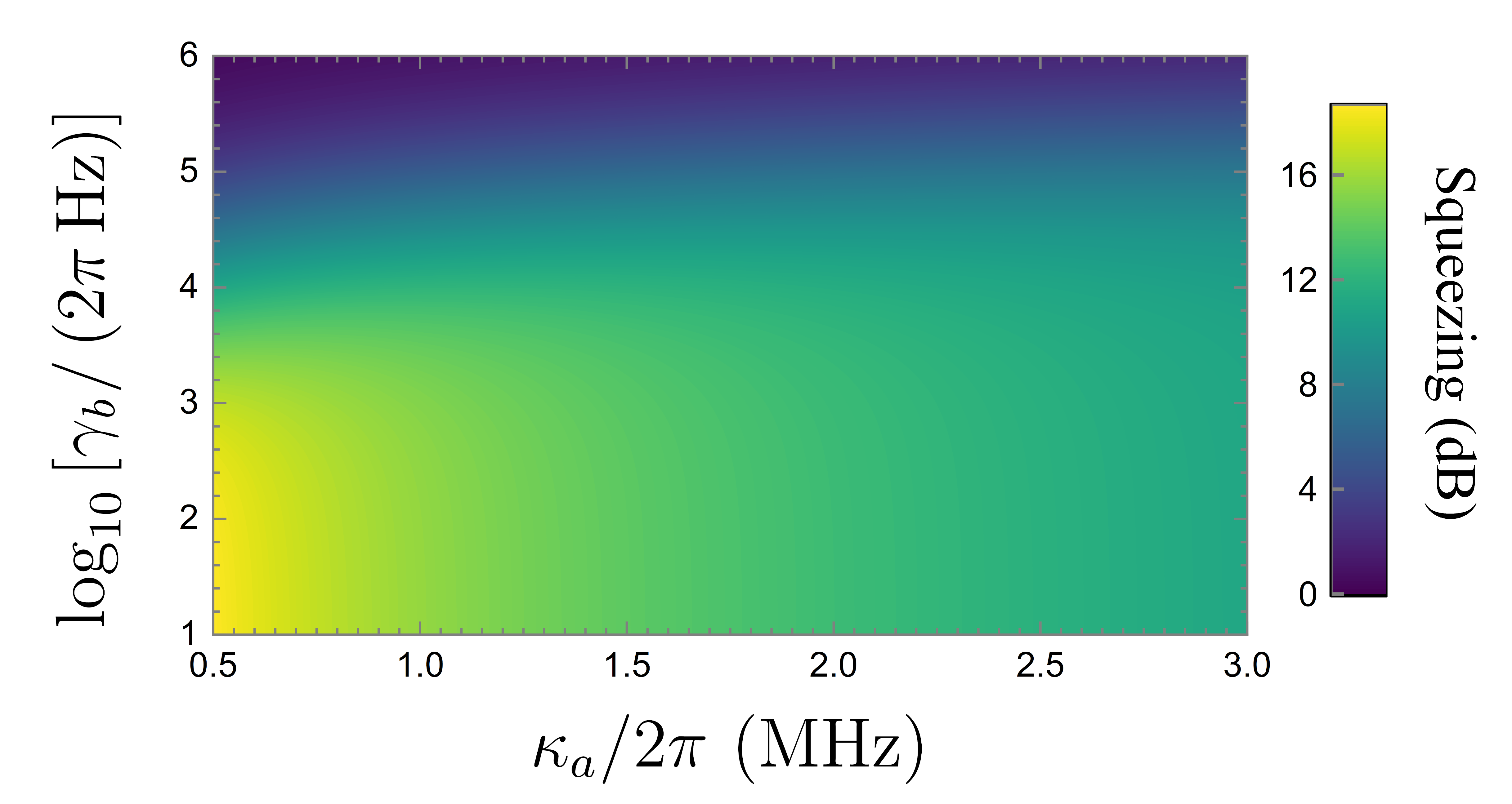}
\caption{The degree of squeezing versus $\kappa_a$ and $\gamma_b$. We take $T=10$ mK, and $G_+$ is optimized for squeezing. The other parameters are the same as those in Fig.~\ref{fig2}(c).}
\label{fig4}
\end{figure}

In Fig.~\ref{fig3}, we plot the squeezing versus the two coupling strengths $G_-$ and $g$. It shows that the squeezing increases with the growth of $G_-$ (with $G_+$ being optimized), and there is an optimal cavity-magnon coupling strength $g$ for getting the maximal squeezing. The scheme is powerful and squeezing below the shot-noise level ($S>0$) can be achieved even for a weak coupling $G_-/2\pi < 0.1$ MHz. 
We further explore the impact of the dissipation rate of each mode on the squeezing in Fig.~\ref{fig4}. Since the magnon decay rate does not change much for the YIG sphere and is about $\kappa_m/2\pi \approx 1$ MHz limited by the intrinsic loss, we plot the squeezing versus the other two decay rates $\gamma_b$ and $\kappa_a$. It reveals that the squeezing is very robust against the mechanical damping, and $S>0$ for a large damping rate $\gamma_b >10^6$ Hz. For the cavity decay rate increasing up to $\kappa_a=2\pi \times 3$ MHz $\ll \omega_b$ allowed by taking a valid RWA, a substantial squeezing greater than 11 dB can still be achieved.

Lastly, we briefly discuss the underlying mechanism of producing output squeezing in the system. The slightly greater coupling $G_->G_+$ associated with the red-detuned drive field eliminates the thermal noise from the system~\cite{RE3,RE4,RE5,RE7,RE8,RE10}, which is the precondition for having the noise reduction below the shot-noise level.  The squeezing mechanism by the combination of the two drive fields can be understood as follows.  The red-detuned drive couples the magnon and mechanical modes through a beam-splitter interaction described by the effective Hamiltonian:
$\hbar G_{-} \big( \delta \hat{b} \delta \hat{m} ^\dagger + \delta \hat{b}^\dagger \hat{m} \big) \equiv \hbar G_{-} \big( \delta \hat{X}_{b} \delta\hat{X}_m + \delta\hat{Y}_b \delta\hat{Y}_m \big)$ [cf. Eq.~\eqref{QLEs_RWA}].  By contrast, the blue-detuned drive couples the two modes via a two-mode squeezing interaction with the effective Hamiltonian:
$\hbar G_{+}  \big( \delta\hat{b}\hat{m} + \delta\hat{b}^\dagger \delta\hat{m}^\dagger \big) \equiv \hbar G_{+} \big( \delta\hat{X}_b \delta\hat{X}_m - \delta \hat{Y}_b\delta\hat{Y}_m \big)$.
Here, $\delta\hat{X}_j$ and $\delta\hat{Y}_j$ ($j=m,b,a$) denote the fluctuations of the two quadratures of each mode. The effective interaction Hamiltonian of the whole system is then given by
\begin{equation}\label{Hamiltonian_new}
\begin{split}
 \hat{H}_{\rm int}/\hbar&=g\left(\delta\hat{X}_a \delta\hat{X}_m + \delta\hat{Y}_a\delta\hat{Y}_m\right)\\
  &\quad\quad\quad \Big(G_-+G_+ \Big)\delta\hat{X}_b \delta\hat{X}_m + \Big(G_--G_+\Big) \delta \hat{Y}_b\delta\hat{Y}_m.
\end{split}
\end{equation}
It clearly shows that the two drive fields lead to unbalanced coupling strengths, i.e., $G_-+G_+$ and $G_--G_+$, between different quadratures of the magnon and mechanical modes, which gives rise to quadrature squeezing of both the magnon and mechanical modes for an initial thermal state. The squeezing is transferred from the magnon mode to the microwave cavity mode, and finally to the cavity output field, as witnessed in Figs.~\ref{fig2}, \ref{fig3}, and \ref{fig4}.

\section{Effect of the CRW terms}
\label{exact}

The results presented in the preceding section are under the RWA, i.e., based on the QLEs~\eqref{QLEs_RWA}, which are a good approximation when $\kappa_{a(m)}, g, G_{\pm} \ll \omega_b$ is satisfied~\cite{RE7}. In this section, we verify this by numerically solving the full QLEs~\eqref{QLEs_fluctuation} and focus on the effect of the CRW terms on the output squeezing, which becomes non-negligible when the above condition is no longer well fulfilled.

We rewrite QLEs~\eqref{QLEs_fluctuation} in the matrix form of
\begin{equation}\label{QLEs_u}
  \dot{u}(t)=A(t)u(t)+n(t),
\end{equation}
where $u(t)=[\delta\hat{a}(t),\, \delta\hat{a}^\dagger(t), \, \delta\hat{b}(t), \, \delta\hat{b}^\dagger(t), \, \delta\hat{m}(t),\,  \delta\hat{m}^\dagger(t)]^{\rm T}$, $n(t)=[\sqrt{\kappa_a} \hat{a}_{in}(t), \sqrt{\kappa_a} \hat{a}_{in}^\dagger (t), \sqrt{\gamma_b} \hat{b}_{in} (t), \sqrt{\gamma_b} \hat{b}_{in}^\dagger (t), \sqrt{\kappa_m} \hat{m}_{in} (t)$, $\sqrt{\kappa_m} \hat{m}_{in}^\dagger (t)]^{\rm T}$, and the {\it time-dependent} drift matrix $A(t)$ is given by
\begin{widetext}
\begin{equation}\label{At}
  A(t) = \begin{pmatrix}
           -\frac{\kappa_a}{2} & 0 & 0 & 0 & -ig & 0 \\
           0 & -\frac{\kappa_a}{2} & 0 & 0 & 0 & ig \\
           0 & 0 & -\frac{\gamma_b}{2} & 0 & -i\left(G_-+e^{2i\omega_bt}G_+\right) & -i\left(G_++e^{2i\omega_bt}G_-\right) \\
           0 & 0 & 0 & -\frac{\gamma_b}{2} & i\left(G_++e^{-2i\omega_bt}G_-\right) & i\left(G_-+e^{-2i\omega_bt}G_+\right) \\
           -ig & 0 & -i\left(G_-+e^{-2i\omega_bt}G_+\right) & -i\left(G_++e^{2i\omega_bt}G_-\right) & -\frac{\kappa_m}{2} & 0 \\
           0 & ig & i\left(G_++e^{-2i\omega_bt}G_-\right) & i\left(G_-+e^{2i\omega_bt}G_+\right) & 0 & -\frac{\kappa_m}{2}
         \end{pmatrix}.
\end{equation}
\end{widetext}
Following Ref.~\cite{NJP}, we express the above drift matrix as $A(t) =A_0 + A_1 e^{2i\omega_bt} + A_{-1} e^{-2i\omega_bt}$, where $A_0$, $A_1$ and $A_{-1}$ are time-independent matrices, given by
\begin{widetext}
\begin{equation}\label{A0}
  A_0 = \begin{pmatrix}
           -\frac{\kappa_a}{2} & 0 & 0 & 0 & -ig & 0 \\
           0 & -\frac{\kappa_a}{2} & 0 & 0 & 0 & ig \\
           0 & 0 & -\frac{\gamma_b}{2} & 0 & -iG_- & -iG_+ \\
           0 & 0 & 0 & -\frac{\gamma_b}{2} & iG_+ & iG_- \\
           -ig & 0 & -iG_- & -iG_+ & -\frac{\kappa_m}{2} & 0 \\
           0 & ig & iG_+ & iG_- & 0 & -\frac{\kappa_m}{2}
         \end{pmatrix},\quad
  A_1=\begin{pmatrix}
        0 & 0 & 0 & 0 & 0 & 0 \\
        0 & 0 & 0 & 0 & 0 & 0 \\
        0 & 0 & 0 & 0 & -iG_+ & -iG_- \\
        0 & 0 & 0 & 0 & 0 & 0 \\
        0 & 0 & 0 & -iG_- & 0 & 0 \\
        0 & 0 & 0 & iG_+ & 0 & 0
      \end{pmatrix},\quad
  A_{-1}=\begin{pmatrix}
        0 & 0 & 0 & 0 & 0 & 0 \\
        0 & 0 & 0 & 0 & 0 & 0 \\
        0 & 0 & 0 & 0 & 0 & 0 \\
        0 & 0 & 0 & 0 & iG_- & iG_+ \\
        0 & 0 & -iG_+ & 0 & 0 & 0 \\
        0 & 0 & iG_- & 0 & 0 & 0
      \end{pmatrix}.
\end{equation}
\end{widetext}
Equation~\eqref{QLEs_u} can then be written in the form of $\dot{u}(t)=$ $\left(A_0 + A_1e^{2i\omega_bt} + A_{-1}e^{-2i\omega_bt} \right)$ $u(t)$ $ + n(t)$. Working in the frequency space by taking the Fourier transform of the equation, we obtain
\begin{equation}
-i\omega u[\omega]=A_0 u[\omega] + A_1 u[\omega+2\omega_b] +A_{-1} u[\omega-2\omega_b]+n[\omega].
\end{equation}
This equation involves three frequency components: $u[\omega]$ and $u[\omega \pm 2\omega_b]$. By replacing $\omega$ with $\omega\pm2k \omega_b$ ($k=1,2,..., l$), we obtain a series of equations involving frequency components $u[\omega\pm2s\omega_b]$ ($s=0,1,2,..., l+1$). These equations can be cast in the following matrix form:
\begin{widetext}
\begin{equation}\label{Fmethods}
  -i\omega\begin{pmatrix}
            ... \\
            u[\omega+4\omega_b] \\
            u[\omega+2\omega_b] \\
            u[\omega] \\
            u[\omega-2\omega_b] \\
            u[\omega-4\omega_b] \\
            ...
          \end{pmatrix} = \begin{pmatrix}
                            ... & ... & ... & ... & ... & ... & ... \\
                            ... & A_0+4i\omega_bI_6 & A_{-1} & 0 & 0 & 0 & ... \\
                            ... & A_1 & A_0+2i\omega_bI_6 & A_{-1} & 0 & 0 & ... \\
                            ... & 0 & A_1 & A_0 & A_{-1} & 0 & ... \\
                            ... & 0 & 0 & A_1 & A_0-2i\omega_bI_6 & A_-1 & ... \\
                            ... & 0 & 0 & 0 & A_1 & A_0-4i\omega_bI_6 & ... \\
                            ... & ... & ... & ... & ... & ... & ...
                          \end{pmatrix}\begin{pmatrix}
            ... \\
            u[\omega+4\omega_b] \\
            u[\omega+2\omega_b] \\
            u[\omega] \\
            u[\omega-2\omega_b] \\
            u[\omega-4\omega_b] \\
            ...
          \end{pmatrix}+\begin{pmatrix}
            ... \\
            n[\omega+4\omega_b] \\
            n[\omega+2\omega_b] \\
            n[\omega] \\
            n[\omega-2\omega_b] \\
            n[\omega-4\omega_b] \\
            ...
          \end{pmatrix},
\end{equation}
\end{widetext}
where $I_6$ denotes a $6\times6$ identity matrix. By solving Eq.~\eqref{Fmethods}, we can obtain the solution of $u[\omega]$ (specifically $a[\omega]$), based on which the NSD of the cavity output field $S_{\phi}^{\rm out} [\omega]$ can be achieved.

\begin{figure}[t]
\hskip-0.5cm\includegraphics[width=0.9\linewidth]{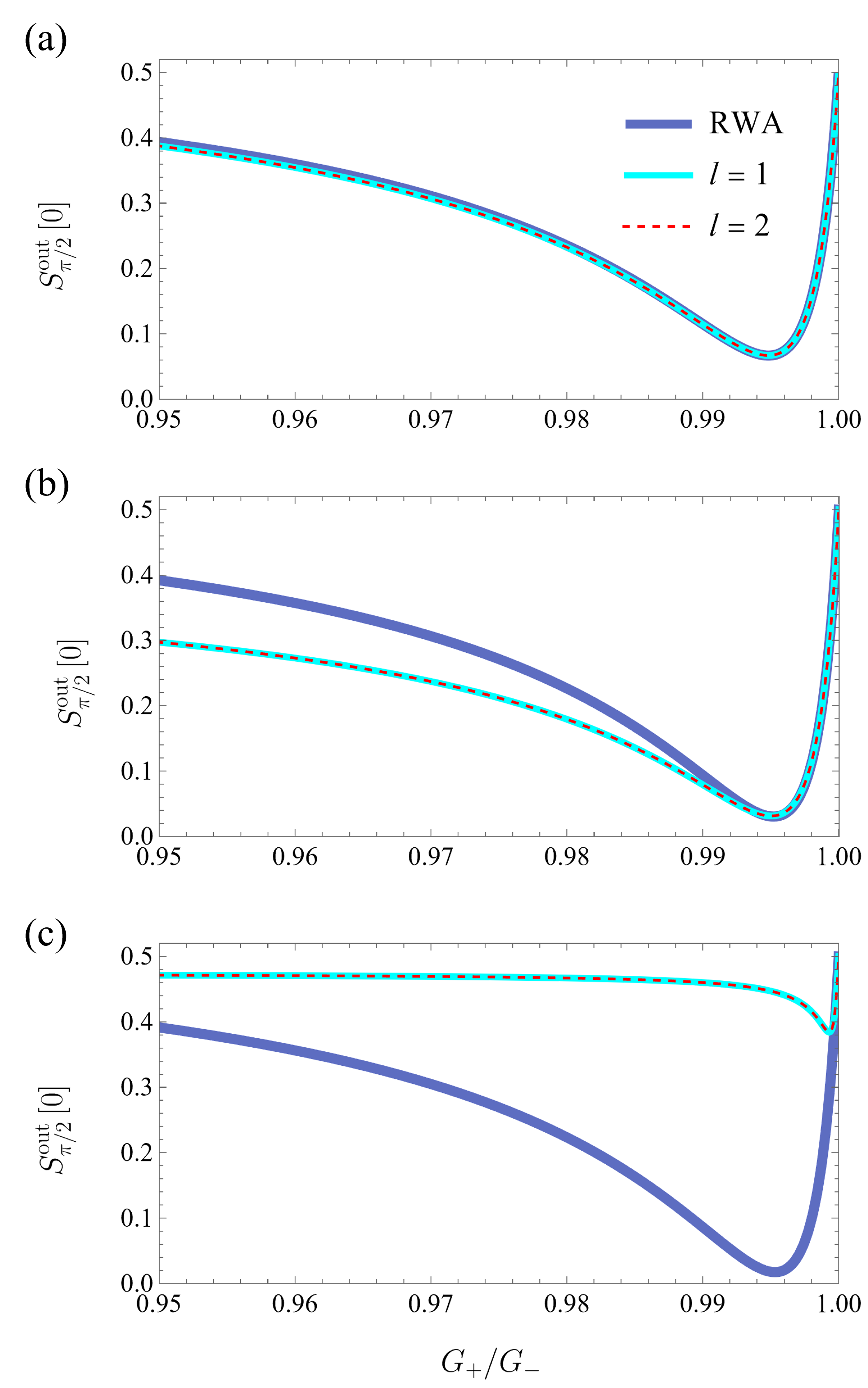}
\caption{The NSD of the output field versus $G_+/G_-$ for $g = G_- =0.05\omega_b$, $0.1\omega_b$, and $0.5\omega_b$ in (a), (b), and (c), respectively. We take $T=10$ mK, and the other parameters are the same as those in Fig.~\ref{fig2}(c).}
\label{fig5}
\end{figure}

In Fig.~\ref{fig5},  we plot $S_{\phi}^{\rm out} [0]$ at the optimal phase angle $\phi=\frac{\pi}{2}$ versus $G_+/G_-$. The thick solid curve is obtained using the analytical result of Eq.~\eqref{Sout} under the RWA and the thin solid (dashed) curve is obtained numerically by solving Eq.~\eqref{Fmethods} and truncating $k$ up to $l=1$ ($l=2$). It reveals that when the condition of $\kappa_{a(m)}, g, G_{\pm} \ll \omega_b$ is satisfied, the results with the RWA are a good approximation, especially around the optimal ratio $G_+/G_-$ for the squeezing, cf. Figs.~\ref{fig5}(a) and \ref{fig5}(b). When the couplings $g$ and $G_{\pm}$ become stronger and comparable to the mechanical frequency $\omega_b$, e.g., $g = G_- =0.5\omega_b$ in Fig.~\ref{fig5}(c), the RWA is no longer valid and the CRW terms play a significant role in reducing the squeezing. Consequently, a large deviation exists between the results with and without taking the RWA. The fact that the thin solid curve and the dashed curve completely overlap in all the plots of Fig.~\ref{fig5} indicates that truncating $k$ up to $l=1$ is already a good approximation and the results are very close to the exact solution.


\section{Conclusion}\label{conc}

We present a protocol for generating squeezed microwave output fields using a cavity magnomechanical system. This is achieved by simultaneously driving the magnon mode with two red- and blue-detuned microwave fields, which induce the magnomechanical beam-splitter and two-mode squeezing interactions, respectively. The simultaneous presence of these two interactions leads to the cooling and squeezing of the mechanical and magnon modes.  The cavity mode then gets squeezed due to the cavity-magnon state swapping, thus yielding a squeezed cavity output field.   Significant squeezing of the microwave output field can be achieved, which is stationary and robust against bath temperature. It is worth noting that the protocol is not limited to YIG spheres, but applicable to any cavity magnomechanical systems that exhibit a dispersive magnon-phonon coupling. The work would find potential applications in quantum information science and quantum metrology.

\section*{Acknowledgments}

This work has been supported by National Key Research and Development Program of China (Grant No. 2022YFA1405200) and National Natural Science Foundation of China (Grant No. 92265202).

\section*{Appendix}
\label{appendix}

\setcounter{figure}{0}
\setcounter{equation}{0}
\setcounter{table}{0}
\renewcommand\theequation{A\arabic{equation}}
\renewcommand\thefigure{A\arabic{figure}}
\renewcommand\thetable{A\arabic{table}}

\begin{figure}[b]
\hskip-0.2cm\includegraphics[width=0.85\linewidth]{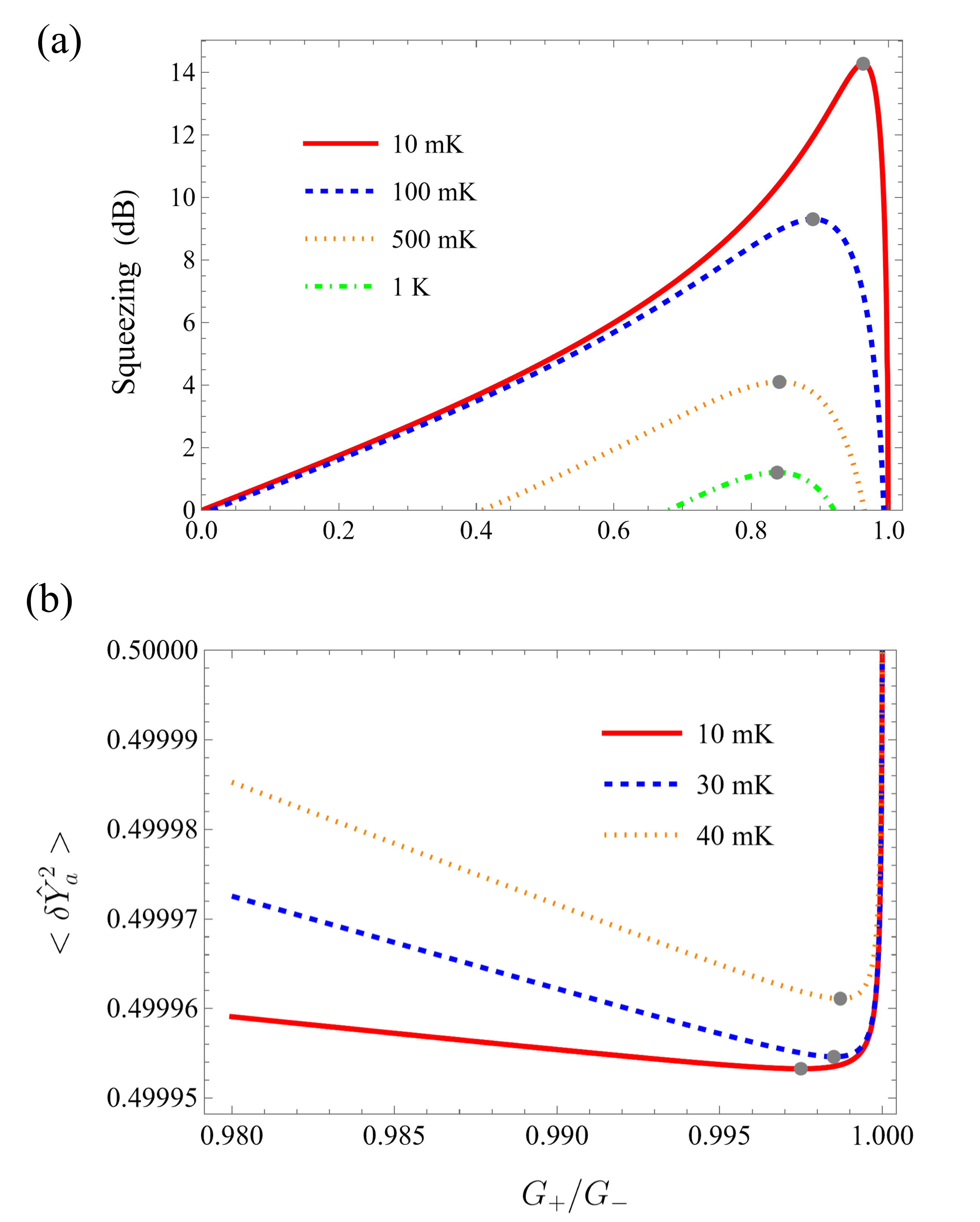}
\caption{(a) The degree of squeezing (dB) of the mechanical position ($\langle \delta \hat{X}_b^2\rangle <\frac{1}{2}$); and (b) Variance of the phase quadrature of the cavity field $\langle \delta \hat{Y}_a^2\rangle$ versus $G_+/G_-$ for various bath temperatures. The gray points correspond to the  optimal squeezing at different temperatures. The other parameters are the same as in Fig.~\ref{fig2}(b).}
\label{figA}
\end{figure}

Here we present the results of the stationary squeezing of the mechanical and cavity modes. The squeezing is evaluated based on the steady-state covariance matrix (CM) of the system, which can be directly obtained by solving the Lyapunov equation, as detailed below.

Specifically, we rewrite the QLEs \eqref{QLEs_RWA} under the RWA in the matrix form of
\begin{equation}\label{QLES_quadratures}
  \dot{u'}(t)=A'u'(t)+n'(t),
\end{equation}
where $u'(t)=[\delta \hat X_a(t),\delta \hat Y_a(t),\delta \hat X_b(t), \delta \hat Y_b(t), \delta \hat X_m(t), \delta \hat Y_m(t)]^{\rm T}$ is the vector of the quadrature fluctuations of the system, $n'(t)=[\!\!\sqrt{\kappa_a} \hat X_a^{in}(t), \!\! \sqrt{\kappa_a} \hat Y_a^{in}(t), \!\! \sqrt{\gamma_b} \hat X_b^{in}(t), \!\! \sqrt{\gamma_b} \hat Y_b^{in}(t), \!\! \sqrt{\kappa_m} \hat X_m^{in}(t), \!\! \sqrt{\kappa_m} \hat Y_m^{in}(t)]^{\rm T}$, and the drift matrix $A'$ is given by
\begin{equation}\label{drift_matrix}
  A'=\begin{pmatrix}
    -\frac{\kappa_a}{2} & 0 & 0 & 0 & 0 & g \\
    0 & -\frac{\kappa_a}{2} & 0 & 0 & -g & 0 \\
    0 & 0 & -\frac{\gamma_b}{2} & 0 & 0 & G_-{-}G_+ \\
    0 & 0 & 0 & -\frac{\gamma_b}{2} & -\big(G_-{+}G_+\big) & 0 \\
    0 & g & 0 & G_-{-}G_+ & -\frac{\kappa_m}{2} & 0 \\
    -g & 0 & -\big(G_-{+}G_+\big) & 0 & 0 & -\frac{\kappa_m}{2}
  \end{pmatrix}.
\end{equation}
The steady state of the quantum fluctuations of the system is a three-mode Gaussian state, which is fully characterized by a $6\times6$ CM $V$, defined as $V_{ij} = \frac{1}{2}\langle u'_i(t)u'_j(t')+u'_j(t')u'_i(t)\rangle$ $(i,j=1,2,...,6)$. The steady-state $V$ can be achieved by solving the Lyapunov equation~\cite{Jie18}:
\begin{equation}\label{Lya}
  A'V+VA'^{\rm T}=-D',
\end{equation}
where the diffusion matrix $D'$ is defined as $\langle n'_i(t)n'_j(t') + n'_j(t') n'_i(t) \rangle/2=D'_{ij}\delta(t-t')$, and it takes the form of $D'={\rm diag}[\kappa_a(N_a+1/2), \kappa_a(N_a+1/2), \gamma_b(N_b+1/2), \gamma_b(N_b+1/2), \kappa_m(N_m+1/2), \kappa_m(N_m+1/2)]$. For each mode, squeezing means that one of the two quadratures of the mode has a variance smaller than the shot-noise level, i.e., $\langle \delta \hat X_j^2\rangle$ or  $\langle \delta \hat Y_j^2\rangle < \frac{1}{2}$ ($j=a,b,m$).  $\langle \delta \hat X_j^2\rangle$ and $\langle \delta \hat Y_j^2\rangle $ can be extracted from the diagonal elements of $V$.

In Fig.~\ref{figA}, we show the squeezing of the corresponding quadrature of the mechanical and cavity modes versus $G_+/G_-$ for different bath temperatures.  Clearly, as the temperature rises, the optimal $G_+/G_-$ decreases for the mechanical squeezing, but it shows an opposite trend for the cavity squeezing, which is consistent with that for the output field squeezing [cf. Fig. \ref{fig2}(b)]. Note that here the cavity squeezing is small because we obtain the steady-state squeezing, and the dynamical squeezing can be much stronger ~\cite{HFW}.

\end{document}